\documentclass[sigconf]{acmart}
\pdfoutput=1 
\renewcommand\footnotetextcopyrightpermission[1]{} 
\usepackage{booktabs} 

\usepackage{multirow}

\acmConference[FAT/ML]{}{4th Workshop on Fairness, Accountability, \\ and Transparency in Machine Learning}{2017} 

\definecolor{myblue}{RGB}{0,30,180}
\definecolor{mygreen}{RGB}{10,120,40}

\begin{document}
\title{Racial Disparity in Natural Language Processing: \\
A Case Study of Social Media African-American English}

\author{Su Lin Blodgett}
\affiliation{%
  \institution{University of Massachusetts Amherst}
  \city{Amherst} 
  \state{MA}
}
\email{blodgett@cs.umass.edu}

\author{Brendan O'Connor}

\affiliation{%
  \institution{University of Massachusetts Amherst}
  \city{Amherst} 
  \state{MA}
}
\email{brenocon@cs.umass.edu}


\begin{abstract}
We highlight an important frontier in algorithmic fairness:
disparity in the quality of natural language processing algorithms
when applied to language from authors of different social groups.
For example, current systems sometimes analyze the language of females and minorities
more poorly than they do of whites and males.
We conduct an empirical analysis of racial disparity in language identification 
for tweets
written in African-American English, and discuss implications of disparity in NLP.
\end{abstract}

%
%



\maketitle

\section{Introduction: Disparity in NLP}

As machine learned algorithms govern more and more 
real-world outcomes, how to make them fair---and what that should mean---is of increasing concern.
One strand of research, heavily represented at the FAT-ML series of workshops,\footnote{\url{http://www.fatml.org/}}
considers scenarios where a learning algorithm must make decisions about people, such as approving prospective applicants for employment, or deciding who should be the targets of police actions \citep{Goel2017Police}, and seeks to develop learners or algorithms
whose decisions have only small differences in behavior between persons from different groups 
\citep{Feldman2015Disparate} or that satisfy other notions of fairness (e.g.~\citep{Joseph2016Rawlsian,Joseph2016Bandits}).

Another recent strand of research has examined a complementary aspect of bias and fairness:
\emph{disparate accuracy in language analysis}.  
Linguistic production
is a critically important form of human behavior, and 
a major class of artificial intelligence algorithms---natural language processing, or language technologies---may or may not fairly analyze language
produced by different types
of authors \citep{hovy2016social}.
For example, Tatman~\citep{tatman:2017:EthNLP} finds that 
YouTube autocaptioning has a higher word error rate
for female speakers than for male speakers in videos.
This has implications for downstream uses of language technology:
\begin{itemize}
\item Viewing: users who rely on autocaptioning have a harder time understanding what women are saying in videos, relative to what men are saying.
\item Access: search systems are necessary for people to access information online, and for videos they
may depend on indexing text recognized from the audio.
Tatman's results \cite{tatman:2017:EthNLP}
imply that such a search system will fail to find information produced by female speakers more often than for male speakers.
\end{itemize}
This bias affects interests of the speakers---it is more difficult for their voices to be communicated to the world---as well as other users, who are deprived of information or opinions from females, or more generally,
any social group whose language experiences lower accuracy of analysis by language technologies.

Gender and dialect are well-known confounds in speech recognition, since they can implicate pitch, timbre, and the pronunciation of words (the phonetic level of language);
domain adaptation is always a challenge and 
research continues on how to apply domain transfer
to speech recognizers across dialects \citep{Lehr2014AAE}.
And more broadly, decades of research in 
the field of \emph{sociolinguistics} has documented an extensive array of both social factors that affect how people produce language (e.g.\ community, geography, ethnicity), and how specifically language is affected (e.g.\ the lexicon, syntax, semantics).  
We might expect
 a minority teenager in school 
 as well as a white middle-aged software engineer
 to both speak English,
but they may exhibit variation in their pronunciation,
word choice, slang, or even syntactic structures.
Dialect communities often align with geographic and sociological factors, as language variation emerges within distinct social networks, or is affirmed as a marker of social identity.

Dialects pose a challenge to fairness in NLP, because they entail language variation that is
\emph{correlated} to social factors, and we believe there needs to be
greater awareness of dialects among technologists using and building language technologies.
In the rest of this paper, we focus on the dialect of African-American English as used on Twitter, which previous work \citep{Blodgett2016AAE,jones2015toward,jorgensen2015challenges}
has established is very prevalent and sometimes quite different than mainstream American English.
We analyze an African-American English Twitter corpus (from Blodgett et al.\ \citep{Blodgett2016AAE}, described in \S\ref{s:model}), and analyze racial disparity in language identification, a crucial first step in any NLP application. Our previous work found that off-the-shelf tools display racial disparity---they tend to erroneously classify messages from African-Americans as non-English more often than those from whites. We extend 
this analysis from 200 to 20,000 tweets, finding that the disparity persists when controlling for 
message length (\S\ref{s:biasexper}), and evaluate the racial disparity for several black-box commercial services.
We conclude with a brief discussion (\S\ref{s:disko}).


\section{African-American English and social media} \label{s:dialects}

We focus on language in social media, which is often informal and conversational.
Social media NLP tools may be used for, say,
sentiment analysis applications,
which
seek to measure opinions from 
online communities.  But current NLP tools are typically trained on traditional written sources,
which are quite different from social media language, and even more so from dialectal social media language.
Not only does this imply social media NLP may be of lower accuracy, but since language can vary across social groups, any such measurements may be biased---incorrectly representing ideas and opinions from people who use non-standard language.


Specifically, we investigate dialectal language in publicly available Twitter data, focusing on African-American English (AAE), a dialect of American English spoken by millions of people across the United States \citep{Labov1972AAE,Rickford1999AAE,Green2002AAE}. AAE is a linguistic variety with defined syntactic-semantic, phonological, and lexical features, which have been the subject of a rich body of sociolinguistic literature. In addition to the linguistic characterization, reference to its speakers and their geographical location or speech communities is important, especially in light of the historical development of the dialect. Not all African-Americans speak AAE, and not all speakers of AAE are African-American; nevertheless, speakers of this variety have close ties with specific communities of African-Americans \citep{Green2002AAE}.

The phenomenon of ``BlackTwitter'' has been noted anecdotally; indeed,
African-American and Hispanic minorities were markedly over-represented in the early years of the Twitter service (as well as younger people) relative to their representation in the American general population.\footnote{\url{http://www.pewinternet.org/fact-sheet/social-media/}}
It is easy to find examples of non-Standard American English (SAE) language use, such as:

\begin{enumerate}
\item \emph{he woke af smart af educated af daddy af coconut oil af GOALS AF \& shares food af}
\item \emph{Bored af den my phone finna die!!!}
\end{enumerate}

\noindent
The first example has low punctuation usage (there is an utterance boundary after every ``af''),
but more importantly,
it displays a key syntactic feature of the AAE dialect, a null copula: ``he woke'' would be written, in Standard American English, as ``he is woke'' (meaning, politically aware).  ``af'' is an online-specific term meaning ``as f---.''  The second example displays two more traditional AAE features: ``den'' is a spelling of ``then'' which follows a common phonological transform in AAE (initial ``th'' changing to a ``d'' sound: ``dat,'' ``dis,'' etc. are also common),
and the word ``finna'' is an auxiliary verb, short for ``fixing to,''
which indicates an immediate future tense (``my phone is going to die very soon''); 
it is part of AAE's rich verbal auxiliary system capable of encoding
different temporal semantics than mainstream English \citep{Green2002AAE}.



\section{Demographic Mixed Membership Model for Social Media} \label{s:model}

In order to test racial disparity in social media NLP, \cite{Blodgett2016AAE} collects a large-scale AAE corpus from Twitter, inferring soft demographic labels with a mixed-membership probabilistic model; we use this same corpus and method, briefly repeating the earlier description of the method.
This approach to identifying AAE-like text makes use of the connection between speakers of AAE and African-American neighborhoods; we harvest
a set of messages from Twitter, cross-referenced against U.S.\ Census demographics, and then analyze words against demographics with a mixed-membership probabilistic model. 
The data is a sample of millions of publicly posted geo-located Twitter messages (from the Decahose/Gardenhose stream \citep{morstatter2013sample}),
most of which are sent on mobile phones, by authors in the U.S. in 2013.

For each message, we look up the U.S.\ Census blockgroup geographic area that the message was sent in,
and use race and ethnicity information for each blockgroup from the Census' 2013 American Community Survey, 
defining four covariates: percentages of the population
that are non-Hispanic whites, non-Hispanic blacks, Hispanics (of any race), and (non-Hispanic) Asians.
Finally, for each user $u$, we average the demographic values of all their messages in our dataset into a length-four vector $\pi^{(census)}_u$.


Given this set of messages and author-associated demographics,
we infer 
statistical associations between language and demographics
with a mixed membership probabilistic model. 
It directly associates each of the demographic variables with a topic; i.e.\ a unigram language model over the vocabulary.
The model assumes an author's mixture over the topics tends to be similar to their Census-associated demographic weights, and that every message has its own topic distribution.
This allows for a single author to use different types of language in different messages,
accommodating multidialectal authors.
The message-level topic probabilities $\theta_m$
are drawn from an asymmetric Dirichlet centered on $\pi^{(census)}_u$,
whose scalar concentration parameter $\alpha$ controls whether authors' language is very similar to the demographic prior, or can have some deviation.  A token $t$'s latent topic $z_t$ 
is drawn from $\theta_m$, and the word itself is drawn from $\phi_{z_t}$, the language model for the topic. Thus, the model learns demographically-aligned language models for each demographic category.
Our previous work \cite{Blodgett2016AAE} verifies that its African-American language model
learns linguistic attributes known in the sociolinguistics literature to be characteristic of AAE,
in line with
other work that has also verified the correspondence of geographical AA prevalence to AAE linguistic features on Twitter \cite{Jorgensen2016AAE,Stewart2014AAE}.

This publicly available corpus contains 59.2 million tweets. 
We filter its messages to ones strongly associated with demographic groups;
for example, for each message we infer the posterior proportion of its tokens that came from the African-American language model, which can be high either due to demographic prior, or from a message that uses many words exclusive to the AA language model (topic); these proportions are available in the released corpus. 
When we filter to messages with AA proportion greater than 0.8,
this results in AAE-like text.  We call these \emph{AA-aligned} messages and
 we also select a set of white-aligned messages in the same way.\footnote{While Blodgett et al.\ verified that the AA-aligned tweets contain well-known features of AAE, we hesitate to call these 
 ``AAE'' and ``SAE'' corpora, since technically speaking they are simply demographically correlated language models.  The Census refers to the categories as ``Black or African-American" and ``White"
 (codes B03002E4 and B03002E3 in ACS 2013).
 And, while
Hispanic- and Asian-associated language models of Blodgett et al.'s model 
are also of interest, we focus our analysis here on the African-American and White language models.}

\section{Bias in NLP Tools} \label{s:biasexper}

\subsection{Language identification}

Language identification, the task of classifying the major world language in which a message is written, is a crucial first step in almost any web or social media text processing pipeline.  For example, in order to analyze the opinions of U.S.\ Twitter users, one might throw away all non-English messages before running an English sentiment analyzer.  (Some of the coauthors of this paper have done this as a simple expedient step in the past.)

A variety of methods for language identification exist \cite{hughes2006reconsidering};
social media 
language identification is particularly challenging
since messages are short and also use non-standard language \citep{baldwin2013noisy}.
In fact, a popular language identification system, \emph{langid.py} \citep{Lui2012Langid},
classifies
both example messages in \S\ref{s:dialects} as Danish with more than 99.9\% confidence.

We take the perspective that since AAE is a dialect of American English,
it ought to be classified as English for the task of major world language identification.
We hypothesize that if a language identification tool
is trained on standard English data,
it may exhibit disparate performance on AA- versus white-aligned tweets.
In particular, we wish to assess the \emph{racial disparity accuracy difference}:

\begin{equation} p( \text{correct} \mid \text{Wh}) - p(\text{correct} \mid \text{AA}) \label{e:absdiff} \end{equation}

\noindent
From manual inspection of a sample of hundreds of messages,
it appears that nearly all white-aligned and AA-aligned tweets are actually English, so accuracy is the same  as proportion of English predictions by the classifier.
A disparity of 0 indicates a language identifier that is fair across these classes.
(An alternative measure is the ratio of accuracies, corresponding to Feldman et al.'s disparate impact measure\ \cite{Feldman2015Disparate}.)

\subsection{Experiments}

We conduct an evaluation of four different off-the-shelf language identifiers, which are popular and
straightforward for engineers to use when building applications:

\begin{itemize}
	\item \textbf{\emph{langid.py} (software):} One of the most popular open source language identification tools, \emph{langid.py} was originally trained on over 97 languages and evaluated on both traditional corpora and Twitter messages \cite{Lui2012Langid}.
	\item \textbf{IBM Watson (API):} The Watson Developer Cloud's Language Translator service supports language identification of 62 languages.\footnote{\url{https://www.ibm.com/watson/developercloud/doc/language-translator/index.html}}
	\item \textbf{Microsoft Azure (API):} Microsoft Azure's Cognitive Services supports language identification of 120 languages.\footnote{\url{https://docs.microsoft.com/en-us/azure/cognitive-services/text-analytics/overview\#language-detection}}
	\item \textbf{Twitter (metadata):} The output of Twitter's in-house identifier, whose predictions are included in a tweet's metadata (from 2013, the time of data collection), which Twitter intends to ``help developers more easily work with targeted subsets of Tweet collections.''\footnote{\url{https://blog.twitter.com/developer/en_us/a/2013/introducing-new-metadata-for-tweets.html}}
	\item \textbf{Google (API, excluded):} We attempted to test 
Google's language detection service,\footnote{\url{https://cloud.google.com/translate/docs/detecting-language}} but it returned a server error for every message we gave it to classify.
\end{itemize}

\noindent
We queried the remote API systems in May 2017.

From manual inspection, we observed that longer tweets are significantly more likely to be correctly classified,
which is a potential confound for a race disparity analysis, since the length distribution is different 
for each demographic group.
To minimize this effect in our comparisons, we group messages into four bins (shown in Table \ref{t:results}) according to the number of words in the message.
For each bin, we sampled 2,500 AA-aligned tweets and 2,500 white-aligned tweets, yielding a total of 20,000 messages across the two categories and four bins.\footnote{Due to a data processing error, there are 5 duplicates (19,995 unique tweets); we report on all 20,000 messages for simplicity.}
We limited pre-processing of the messages to fixing of HTML escape characters and removal of URLs, keeping ``noisy" features of social media text such as @-mentions, emojis, and hashtags. We then calculated, for each bin in each category, the number of messages predicted to be in English by each classifier. Accuracy results are shown in Table \ref{t:results}.\footnote{We have made the 20,000 messages publicly available at: \url{http://slanglab.cs.umass.edu/TwitterAAE/}}

\begin{table*}
\centering
\begin{tabular}{|c|c|c|c|c|} \hline
& & AA Acc. & WH Acc. & Diff. \\ \hline
\multirow{4}{*}{\emph{langid.py}} & $t \leq 5$ & 68.0 & 70.8 & 2.8 \\
& $5 < t \leq 10$ & 84.6 & 91.6 & 7.0 \\
& $10 < t \leq 15$ & 93.0 & 98.0 & 5.0 \\ 
& $t > 15$ & 96.2 & 99.8 & 3.6 \\ \hline
\multirow{4}{*}{IBM Watson} & $t \leq 5$ & 62.8 & 77.9 & 15.1 \\ 
& $5 < t \leq 10$ & 91.9 & 95.7 & 3.8 \\ 
& $10 < t \leq 15$ & 96.4 & 99.0 & 2.6 \\
& $t > 15$ & 98.0 & 99.6 & 1.6 \\ \hline
\multirow{4}{*}{Microsoft Azure} & $t \leq 5$ & 87.6 & 94.2 & 6.6 \\
& $5 < t \leq 10$ & 98.5 & 99.6 & 1.1 \\
& $10 < t \leq 15$ & 99.6 & 99.9 & 0.3 \\
& $t > 15$ & 99.5 & 99.9 & 0.4 \\ \hline
\multirow{4}{*}{Twitter} & $t \leq 5$ & 54.0 & 73.7 & 19.7 \\
& $5 < t \leq 10$ & 87.5 & 91.5 & 4.0 \\
& $10 < t \leq 15$ & 95.7 & 96.0 & 0.3 \\ 
& $t > 15$ & 98.5 & 95.1 & -3.0 \\ \hline
\end{tabular}
\caption{Percent of the 2,500 tweets in each bin classified as English by each classifier; \emph{Diff.} is the difference (disparity on an absolute scale) between the classifier accuracy on the AA-aligned and
 white-aligned samples. 
 $t$ is the message length for the bin.
\label{t:results}}
\end{table*}

As predicted, classifier accuracy does increase as message lengths increase; classifier accuracy is generally excellent for all messages containing at least 10 tokens. 
This result agrees with previous work finding short texts to be challenging to classify
(e.g.\ \cite{baldwin2010language}), since there are fewer features (e.g.\ character n-grams)
to give evidence for the language used.\footnote{A reviewer asked if length is used as a feature; 
we know that the open-source \emph{langid.py} system does not (explicitly) use it.}

However, the classifier results display a disparity in performance among messages of similar length;
for all but one length bin under one classifier, accuracy on the white-aligned sample is higher than on the AA-aligned sample. The disparity in performance between AA- and white-aligned messages is greatest when messages are short; the gaps in performance for extremely short messages ranges across classifiers from 6.6\% to 19.7\%. This gap in performance is particularly critical as 41.7\% of all AA-aligned messages in the corpus as a whole have 5 or fewer tokens.\footnote{For most (system,length) combinations,
the accuracy difference is significant under a two-sided t-test ($p<.01$) except for two rows
($t \leq 5$, \emph{langid.py}, $p=.03$) and 
($10 < t \leq 15$, Twitter, $p=0.5$).
Accuracy rate standard errors range from $0.04\%$ to $0.9\%$
($\approx\sqrt{acc(1-acc)/2500}$).}

%



\section{Discussion} \label{s:disko}
Are these disparities substantively significant?  It is easy to see how statistical bias could arise in downstream applications.  For example, consider an analyst trying to look at
 major opinions about a product or political figure, with a sentiment analysis system that only gathers opinions from messages classified as English by Twitter.  For messages length 5 or less, opinions from African-American speakers will be shown to be $1-54.0/73.7=27\%$ less frequent than they really are, relative to white opinions.
Fortunately, the accuracy disparities are often only a few percentage points;
nevertheless, it is important for practitioners to keep potential biases like these in mind.
 
One way forward to create less disparate NLP systems will be to use domain adaptation and other methods to extend algorithms to work on different distributions of data;
for example, our demographic model's predictions can be used to improve a language identifier,
since the demographic language model's posteriors accurately identify some cases of dialectal English \cite{Blodgett2016AAE}.  In the context of speech recognition, Lehr et al.\ \cite{Lehr2014AAE}
pursue a joint modeling approach, learning pronunciation model parameters for AAE and SAE simultaneously.

One important issue may be the limitation of perspective of technologists versus users.
In striking contrast to Twitter's (historically) minority-heavy demographics, major U.S. tech companies
are notorious for their low representation of African-Americans and Hispanics;
for example,
Facebook and Google report only 1\% of their tech employees are African-American,\footnote{\url{https://newsroom.fb.com/news/2016/07/facebook-diversity-update-positive-hiring-trends-show-progress/}
\url{https://www.google.com/diversity/}} 
as opposed to 13.3\% in the overall U.S.\ population,\footnote{\url{https://www.census.gov/quickfacts/table/RHI225215/00}}
and the population of computer science researchers in the U.S. has similarly low minority representation.
It is of course one example of the ever-present challenge of software designers understanding 
how users use their software; in the context of language processing algorithms,
 such understanding must be grounded in an understanding of dialects and sociolinguistics.

\bibliographystyle{ACM-Reference-Format}
\bibliography{brenocon,emnlp2016} 

\end{document}